\documentclass[preprint,showpacs,preprintnumbers,amssymb,aps]{revtex4}
\usepackage{hyperref}
\usepackage{floatrow}
\usepackage{graphicx}
\usepackage{dcolumn}
\usepackage{amsmath}
\usepackage{epsfig}
\usepackage{subfig}
\usepackage{array}
\usepackage{epstopdf} 
\usepackage{xcolor}
\usepackage{bm}

\def\be{\begin{equation}}
\def\ee{\end{equation}}
\def\bea{\begin{eqnarray}}
\def\eea{\end{eqnarray}}
\def\NO{\nonumber}
\def\gev{\mathrm{~GeV}}
\def\fb{\mathrm{~fb}}

\begin{document}

\normalsize

\title{Is the colour-octet mechanism consistent with the double $J/\psi$ production measurement at B-factories?}

\author{Yu Feng$^1$}
\author{Zhan Sun$^2$}
\author{Hong-Fei Zhang$^{1, 3}$}
\affiliation{
\footnotesize
$^{1}$ Department of Physics, School of Biomedical Engineering, Third Military Medical University, Chongqing 400038, China. \\
$^{2}$ School of Science, Guizhou Minzu University, Guiyang 550025, P. R. China. \\
$^{3}$ School of Science, Chongqing University of Posts and Telecommunications, Chongqing, China.
}

\begin{abstract}
Double $J/\psi$ production in $e^+e^-$ collisions involving colour-octet channels are evaluated up to order $\alpha^2\alpha_s^3$.
Having implemented the variation of the parameters ($m_c$, $\mu_r$ and long-distance matrix elements),
we found that the cross sections for producing double $J/\psi$ at B-factories range from $-0.016$fb to $0.245$fb,
which are even much smaller than that via the colour-siglet mechanism.
Accordingly, this result is consistent with the measurement by the Belle and BABAR Collaborations.
\end{abstract}

\pacs{12.38.Bx, 13.60.Le, 13.88.+e, 14.40.Pq}

\maketitle

\section{Introduction}
The phenomenological study on the nonrelativistic QCD (NRQCD) effective theory~\cite{Bodwin:1994jh} is making new progress since the LHC started its running.
Copious data not only provides evidences for the colour-octet (CO) mechanism,
but also indicates challenges to the theory.
In addition to the fact that the $J/\psi$ hadroproduction data can be well reproduced by the theoretical evaluations within the NRQCD framework~\cite{Braaten:1994vv, Ma:2010yw, Butenschoen:2010rq},
$\chi_c$ hadroproduction~\cite{Ma:2010vd, Jia:2014jfa} gives another strong support.
In low transverse momentum ($p_t$) region, even though the factorization might not hold,
the colour-glass-condensate model~\cite{McLerran:1993ni, McLerran:1993ka, McLerran:1994vd} associated with NRQCD~\cite{Kang:2013hta}
did a good job in the description of the $J/\psi$ production in proton-proton and proton-nucleus collisions~\cite{Ma:2014mri, Ma:2015sia}.
Despite all the successes, we can not overlook the challenges it is facing.
The universality of the NRQCD long-distance matrix elements (LDMEs) has not yet been suggested in all the processes.
As an example, the constraint~\cite{Zhang:2009ym} on the CO LDMEs indicated by the QCD next-to-leading order (NLO) study on the $J/\psi$ production at B factories
is apparently below the LDME values obtained through the fit of the $J/\psi$ production data at other colliders~\cite{Ma:2010yw, Butenschoen:2011yh, Gong:2012ug, Bodwin:2014gia}.
The perspectives of the long-standing $J/\psi$ polarization puzzle still have not converged.
Three groups~\cite{Butenschoen:2012px, Chao:2012iv, Gong:2012ug} achieved the calculation of the $J/\psi$ polarization at hadron colliders at QCD NLO;
however, with different LDMEs, their results are complete different from one another.
Recently, the $\eta_c$ hadroproduction was measured by the LHCb Collaboration~\cite{Aaij:2014bga},
which provides another laboratory for the study of NRQCD.
Ref.~\cite{Butenschoen:2014dra} considers it as a challenge to NRQCD,
while Refs.~\cite{Han:2014jya, Zhang:2014ybe} found these data are consistent with the $J/\psi$ hadroproduction measurements.
Further, with the constraint on the LDMEs obtained in Ref.~\cite{Zhang:2014ybe},
Ref.~\cite{Sun:2015pia} discovered some interesting features of the $J/\psi$ polarization,
and found a possibility of understanding the $J/\psi$ polarization within the NRQCD framework.

The $J/\psi$ pair production at B factories is another challenge that NRQCD is facing.
Belle~\cite{Abe:2004ww} and BABAR~\cite{Aubert:2005tj} Collaborations observed the process $e^+e^-\rightarrow J/\psi+$Charmonium,
and found no evidence for the $J/\psi$ pair events,
while the QCD leading order (LO) calculation based on the colour-siglet (CS) mechanism predicted a significant production rate~\cite{Bodwin:2002fk, Bodwin:2002kk}.
This was understood by the QCD NLO corrections~\cite{Gong:2008ce}, which contribute a negative value and cancel the large LO cross sections.
Ref.~\cite{Gong:2008ce} only talked about the CS contributions.
However, the Belle and BABAR measurements actually did not exclude the double $J/\psi$ plus light hadron events.
Both of the experiments measured the $M_{res}$ spectrum,
where $M_{res}$ denotes the invariant mass of all the final states except for the fully reconstructed $J/\psi$.
These distributions exhibited no significant excess in the range of about 300 MeV above the $J/\psi$ mass,
which suggested that the cross section for the $J/\psi$ pair plus light hadron (e.g. $\pi^0$, $\pi^+\pi^-$) associated production is also too small to observe.
To accord with NRQCD, the double $J/\psi$ production cross sections involving the CO channels must not be significant,
which, however, is not manifest.
Although suppressed by the CO LDMEs,
the double $J/\psi$ yield due to the CO mechanism is enhanced by the powers of $\alpha_s/\alpha$, relative to via the CS channels.
As is pointed out in Refs.~\cite{Bodwin:2002fk, Gong:2008ce}, at B factories,
double $c\bar{c}(^3S_1^{[1]})$ can be produced via two virtual photons generated through the $e^+e^-$ annihilation,
and the LO contribution is of order ${\cal O}(\alpha^4\alpha_s^0)$.
In contrast, as illustrated in FIG.(\ref{diagram-a}, \ref{diagram-b}, and \ref{diagram-c}),
the diagrams for the processes
\be
e^+e^-\rightarrow c\bar{c}(m_1)+c\bar{c}(m_2), \label{eqn:process1}
\ee
when $n_1$ and $n_2$ have the opposite charge conjugation,
involve only a single virtual photon,
and the LO contribution is of order ${\cal O}(\alpha^2\alpha_s^2)$.
For double $J/\psi$ production, $m_1$ and $m_2$ have only two possible configurations,
which are $m_1=^3S_1^{[8]}$ and $m_2=^1S_0^{[8]}$, and $m_1=^3S_1^{[8]}$ and $m_2=^3P_J^{[8]}$.
These two processes are suppressed by the CO LDMEs by a factor of $v^8\approx 0.002$,
where $v$ is the typical charm-quark velocity in the charmonium rest frame,
however, enhanced by the coupling constants by a factor of $\alpha_s^2/\alpha^2\approx 1000$,
relative to via the CS channels.
The double $J/\psi$ can also be produced through such kind of processes,
\be
e^+e^-\rightarrow c\bar{c}(n_1)+c\bar{c}(n_2)+g, \label{eqn:process2}
\ee
where $g$ denotes a gluon.
When $n_1=^3S_1^{[1]}$ and $n_2=^3S_1^{[8]}$ (or equivalently $n_1=^3S_1^{[8]}$ and $n_2=^3S_1^{[1]}$),
this kind of processes are enhanced by the coupling constants by a factor of $\alpha_s^3/\alpha^2\approx 200$ and reduced by the CO LDME by a factor of $v^4\approx 0.05$,
relative to the processes involving only the CS channels.
In sum, the processes involving CO states are enhanced by a synthetic factor of about $2\sim 10$,
comparing with the processes considered in Refs.~\cite{Bodwin:2002fk, Gong:2008ce},
the LO cross sections of which is large enough to be observed by Belle and BABAR experiments.
Accordingly, we need to calculate the cross sections for the $J/\psi$ pair production involving the CO channels to see whether NRQCD can endure this paradox.

In this work, we will present a comprehensive study on the double $J/\psi$ production in $e^+e^-$ annihilation involving CO channels up to order $\mathcal{O}(\alpha^2\alpha_s^3)$,
and check whether it is consistent with the meausurements by Belle and BABAR Collaborations.
The $J/\psi$ plus $\chi_c$ production at B factories has already been studied in Refs.~\cite{Liu:2002wq, Wang:2011qg, Dong:2011fb},
and their results do not contradict the double $J/\psi$ measurements by Belle and BABAR Collaborations,
regarding the branching ratios $\mathcal{B}(\chi_{c[0,1,2]}\rightarrow J/\psi)=[1.27\%$, 33.9$\%$, 19.2$\%]$~\cite{Agashe:2014kda}.
In this paper, we do not calculate the $c\bar{c}(^3S_1^{[1]})+c\bar{c}(^3P_J^{[1]})$ production.
We also notice that the $J/\psi$ may come from the $\chi_{cJ}$ feed down, where the $\chi_{cJ}$ can be produced via the $^3S_1^{[8]}$ channel.
By employing the LDMEs obtained in Refs.~\cite{Jia:2014jfa, Sun:2015pia},
$\langle O^{\chi_{c0}}(^3S_1^{[8]})\rangle=2.01\times 10^{-3}\gev^3$ and $\langle O^{J/\psi}(^3S_1^{[8]})\rangle=1.08\times 10^{-2}\gev^3$ in association with the branching ratios listed above,
we find that this contribution is much smaller than that from the $J/\psi$ directly produced through the $^3S_1^{[8]}$ channel.
Similarly, the $J/\psi$ production cross sections via the $\psi(2s)$ feed down is also smaller than that for the directly produced ones.
For this reason, we completely omit the discussions on the feed down contributions from both $\chi_c$ and $\psi(2s)$.

The rest of this paper is organised as follows. In Sec.\ref{cha:2}, we outline the formalism of the calculation.
Sec.\ref{cha:3} presents the numerical results and related discussions, followed by a concluding remark in Sec.\ref{cha:4}.

\section{Double $J/\psi$ production in NRQCD framework} \label{cha:2}
\begin{figure} 
  \begin{center}
  \subfloat[]{\includegraphics[width=5.0cm]{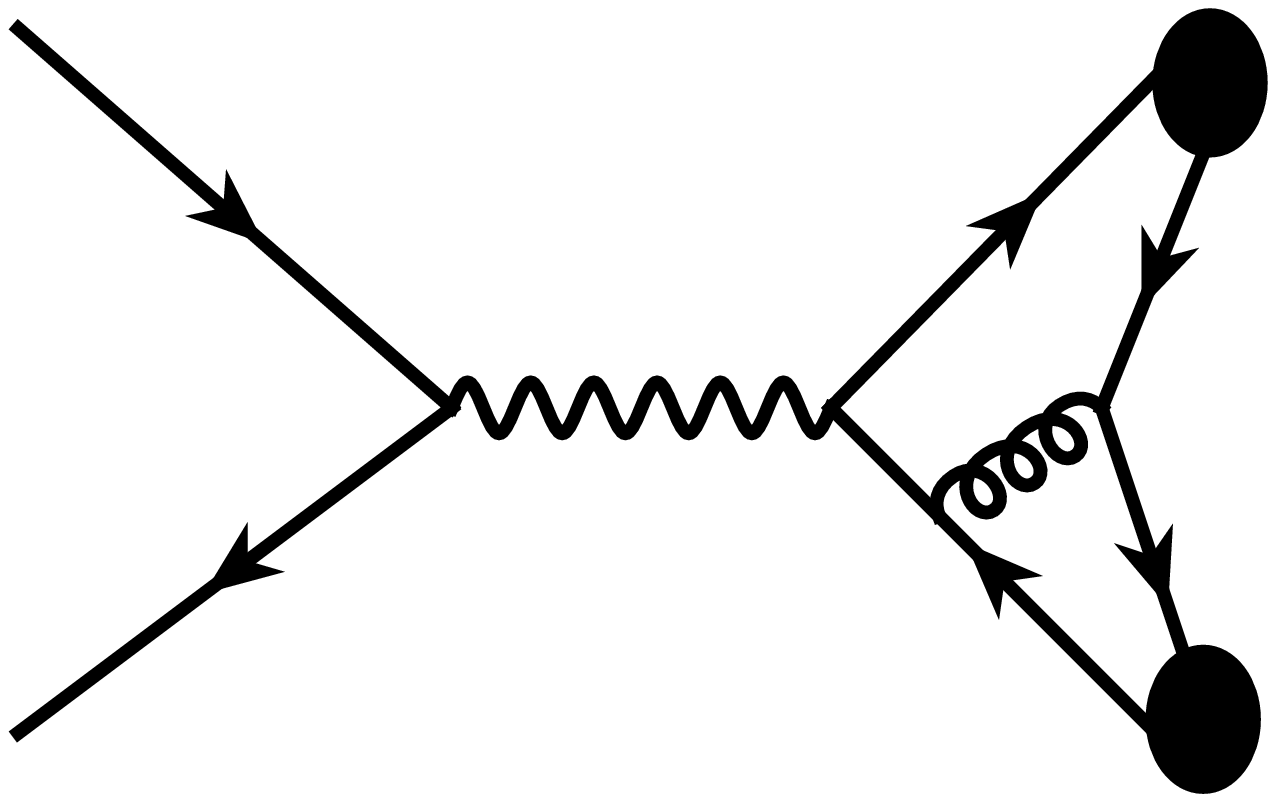} \label{diagram-a}} 
  \subfloat[]{\includegraphics[width=5.0cm]{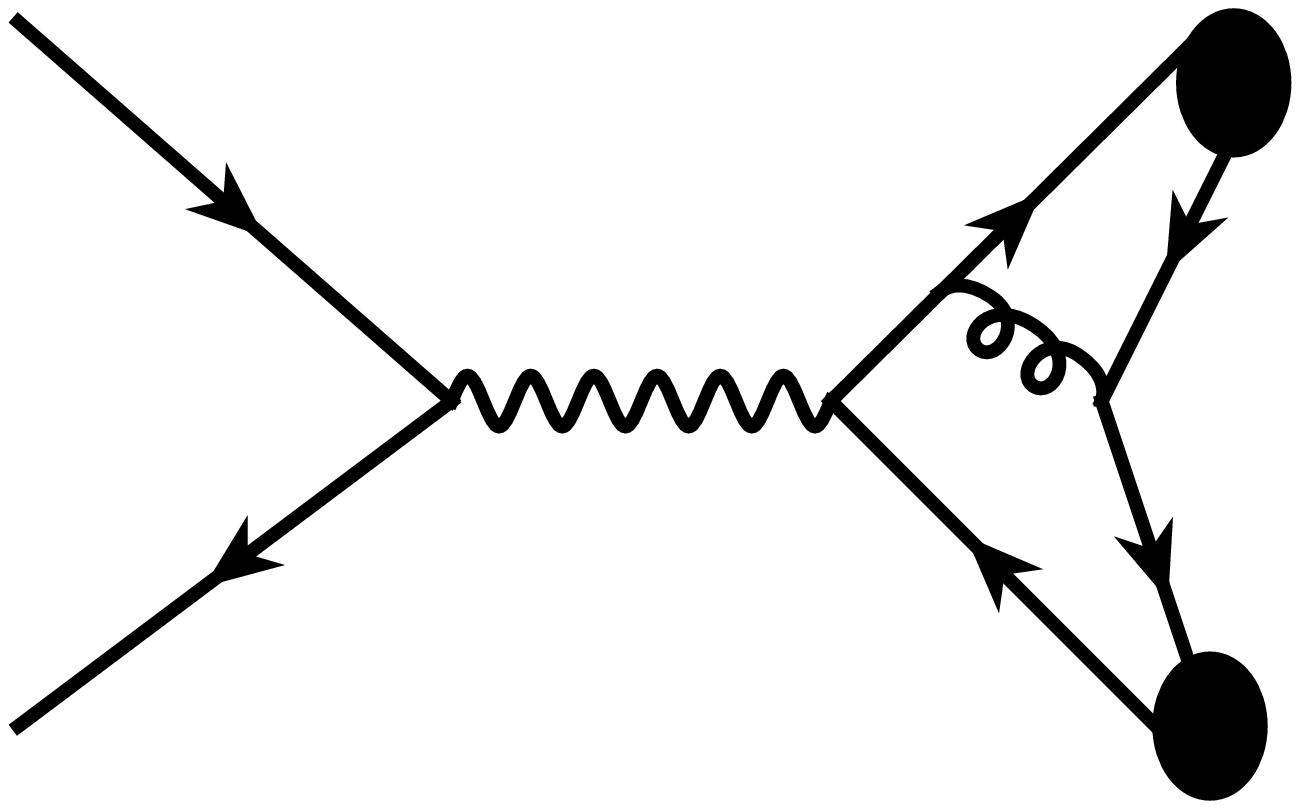} \label{diagram-b}}
  \subfloat[]{\includegraphics[width=5.0cm]{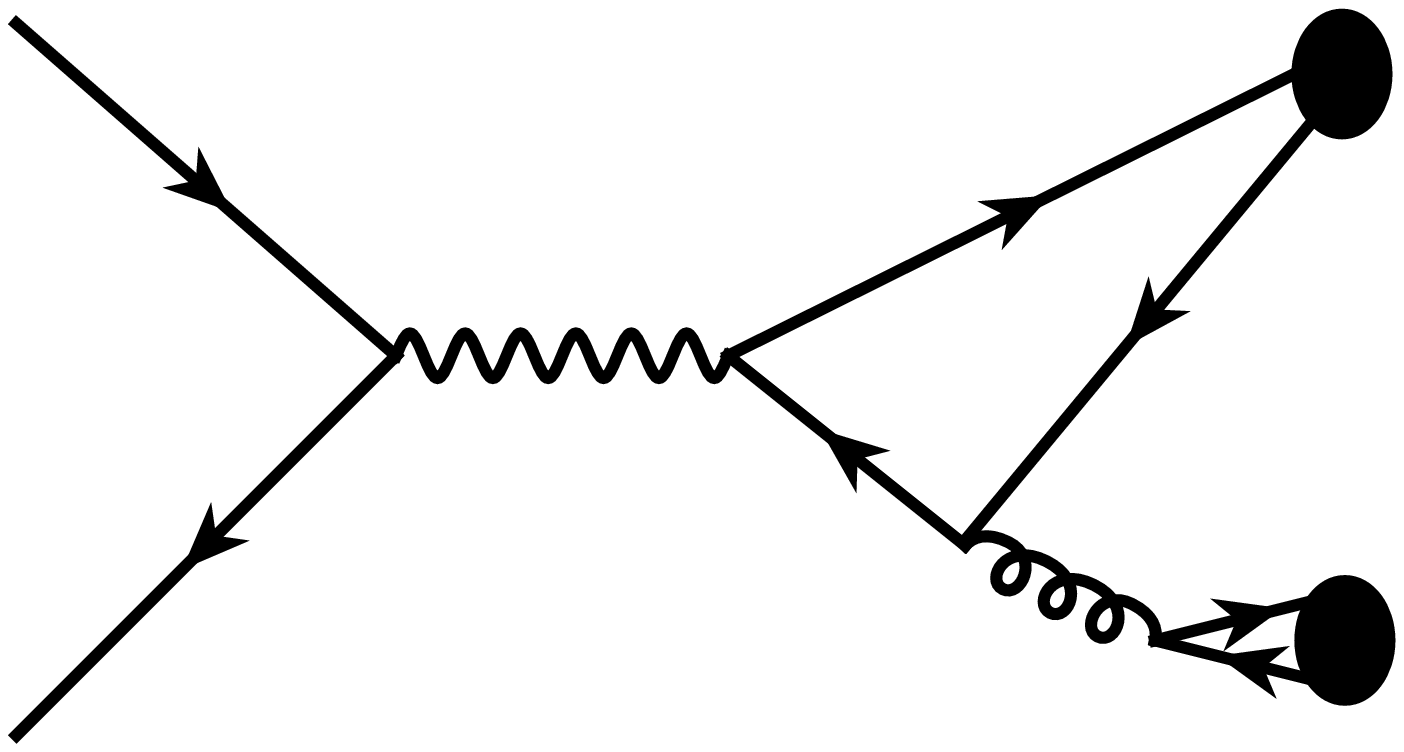} \label{diagram-c}} \\
  \subfloat[]{\includegraphics[width=5.0cm]{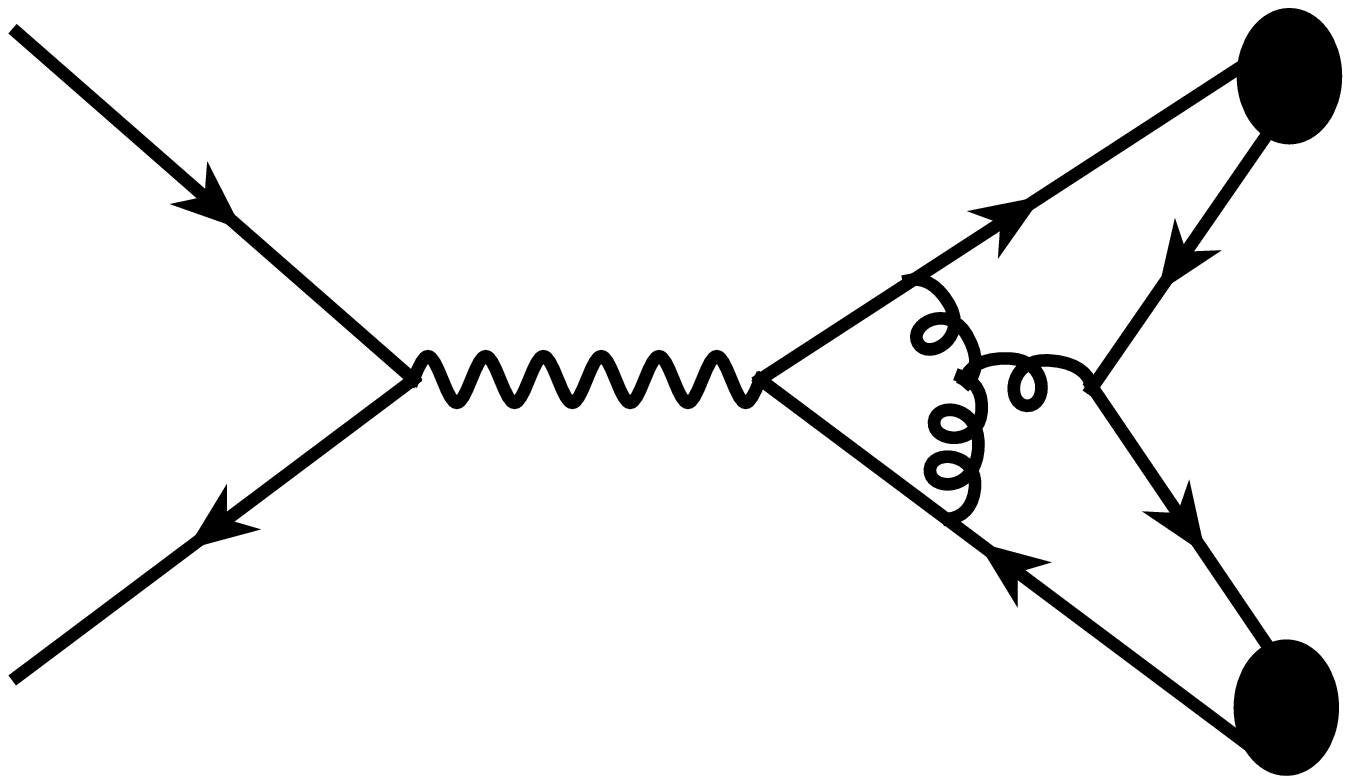} \label{diagram-d}}
  \subfloat[]{\includegraphics[width=5.0cm]{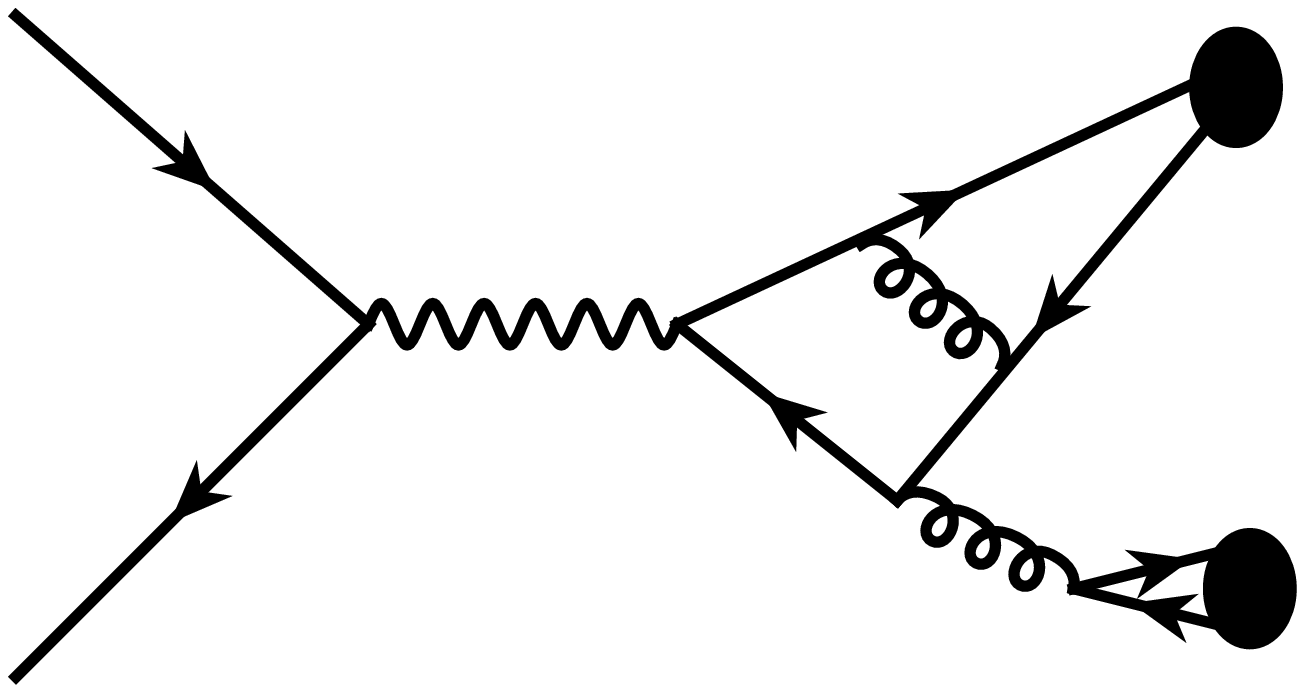} \label{diagram-e}}
  \subfloat[]{\includegraphics[width=5.0cm]{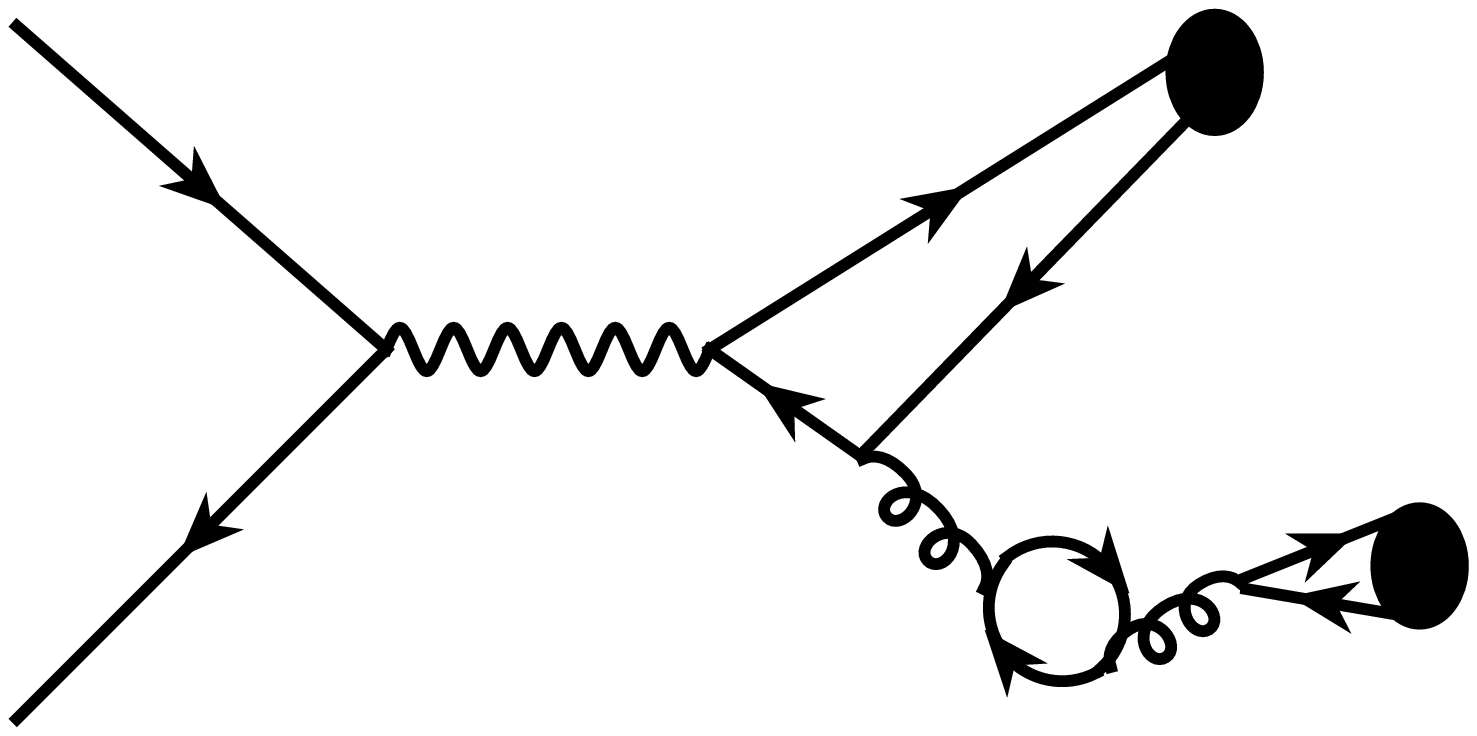} \label{diagram-f}} \\
  \subfloat[]{\includegraphics[width=5.0cm]{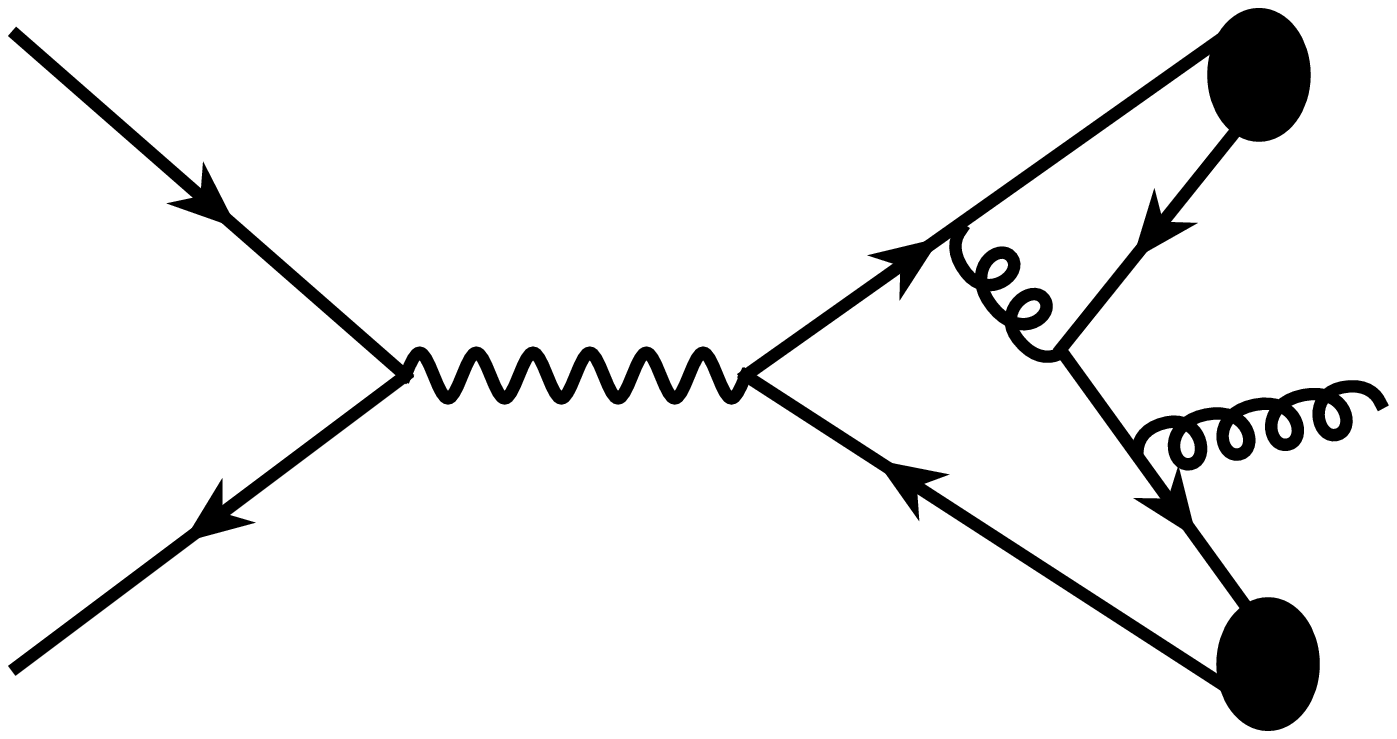} \label{diagram-g}}
  \subfloat[]{\includegraphics[width=5.0cm]{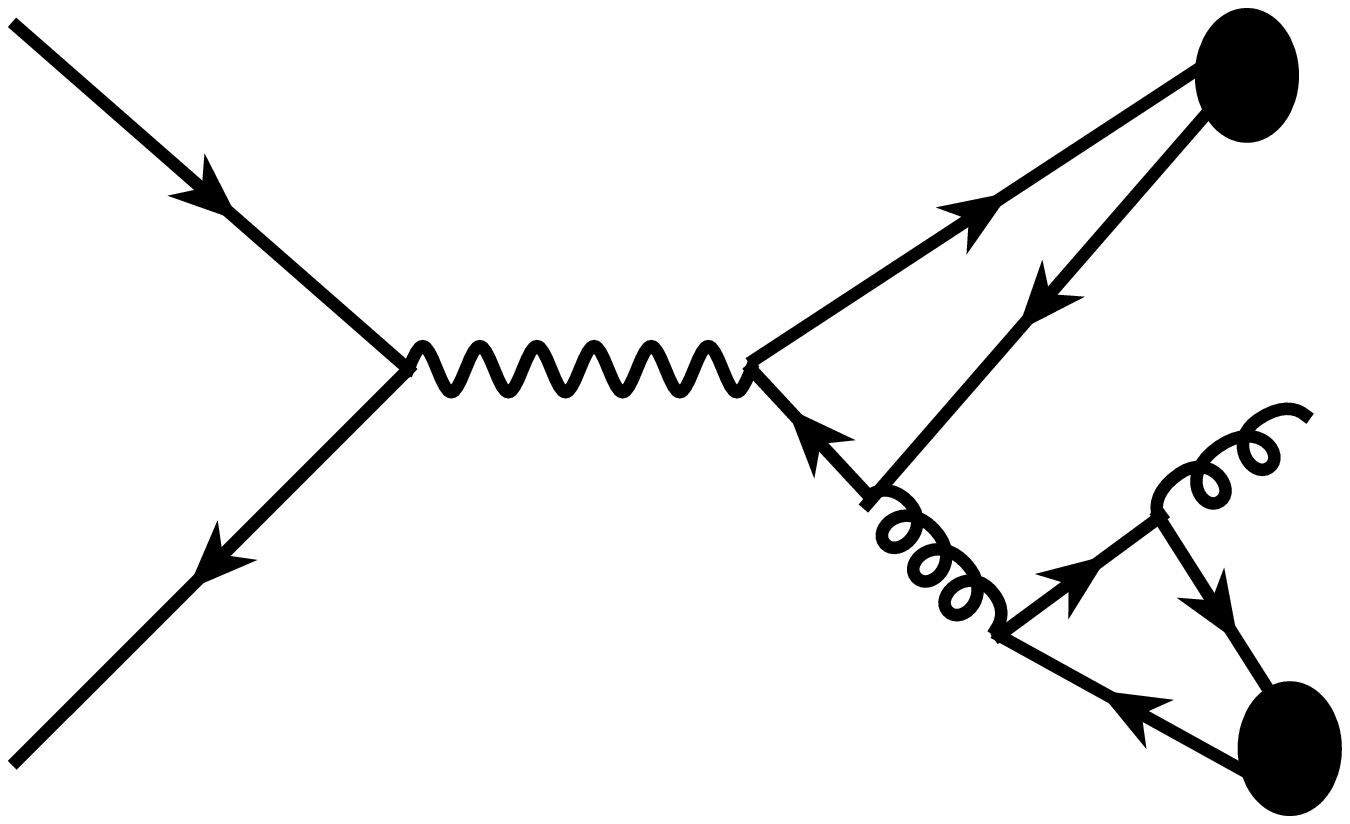} \label{diagram-h}}
  \subfloat[]{\includegraphics[width=5.0cm]{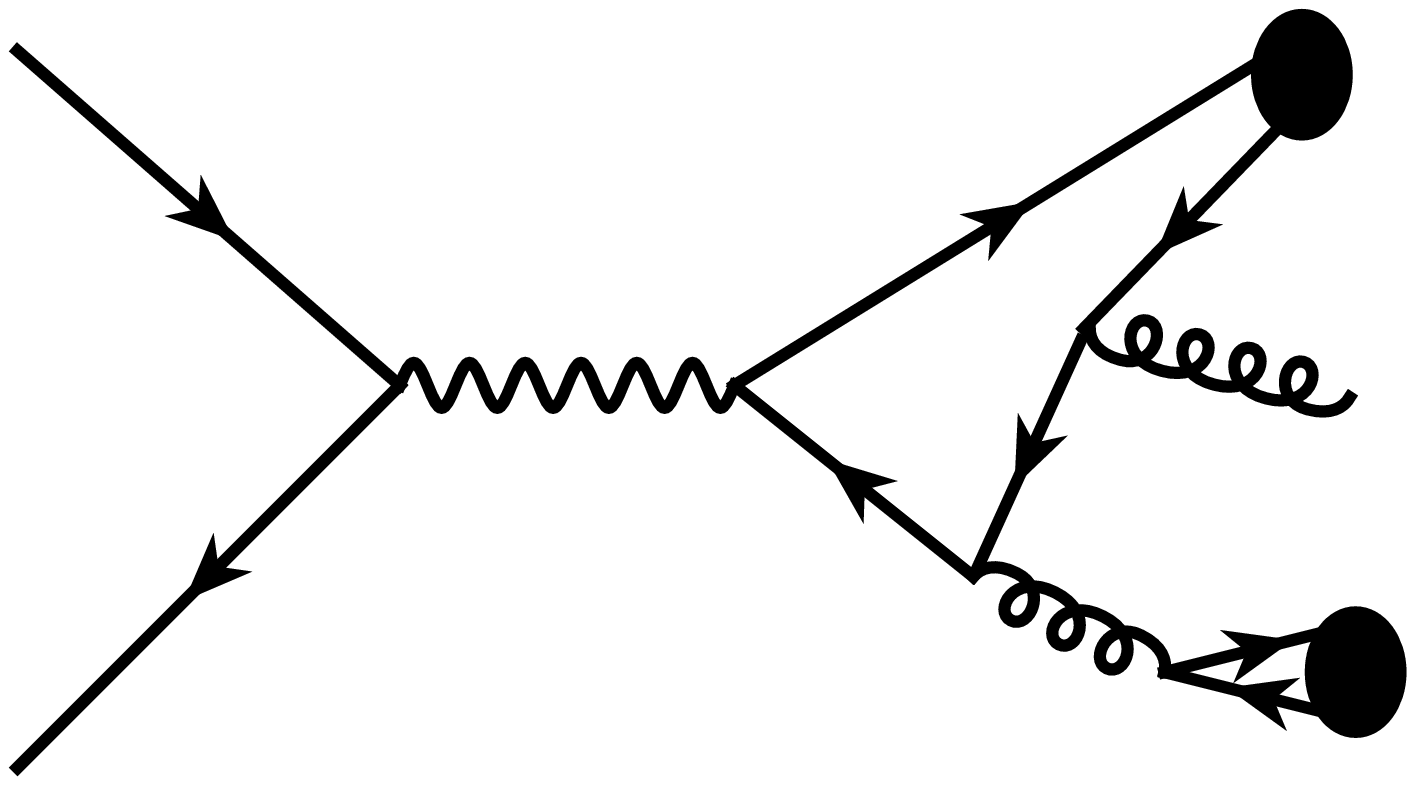} \label{diagram-i}} \\
  \end{center}
  \caption{Representative feynman diagrams.  }
  \label{fig:feyndiag}
\end{figure}

Following the NRQCD factorization, the total cross sections for the $J/\psi$ pair production can be expressed as
\be
\sigma(e^+e^-\rightarrow J/\psi+J/\psi+X)=\sum_{n_1,n_2}\hat{\sigma}(e^+e^-\rightarrow c\overline{c}(n_1)+c\overline{c}(n_2)+X)\langle O^{J/\psi}(n_1)\rangle\langle O^{J/\psi}(n_2)\rangle, \label{eqn:total}
\ee
where $n_1$, $n_2$ run over all the possible configurations of the $c\overline{c}$ intermediate states with certain colour and angular momentum,
$\hat{\sigma}$ is the short-distance coefficient (SDC),
and $\langle O^{J/\psi}(n_1)\rangle$ and $\langle O^{J/\psi}(n_2)\rangle$ are the corresponding LDMEs.
When at least one of $n_1$ and $n_2$ is a CO state, the LO contributions are of order $\mathcal{O}(\alpha_s^2\alpha^2)$.
At this order, all the processes have the form of Eq.(\ref{eqn:process1}),
in which the only possible configurations of $m_1$ and $m_2$ are $m_1=^3S_1^{[8]}$, $m_2=^1S_0^{[8]}$ and $m_1=^3S_1^{[8]}$, $m_2=^3P_J^{[8]}$,
and the representative Feynman diagrams are illustrated in Fig.(\ref{diagram-a}, \ref{diagram-b}, and \ref{diagram-c}).
At QCD NLO ($\mathcal{O}(\alpha_s^3\alpha^2)$), in addition to the virtual corrections
(the representative Feynman diagrams for which are shown in Fig.(\ref{diagram-d}, \ref{diagram-e}, and \ref{diagram-f}))
to the processes presented in Eq.(\ref{eqn:process1}),
double $c\bar{c}$ states in association with a gluon production is also required for consideration, as illustrated in Eq.(\ref{eqn:process2}).
The real-correction processes to the LO ones are
\bea
e^+e^-\rightarrow c\bar{c}(^3S_1^{[8]})+c\bar{c}(^1S_0^{[8]})+g, \NO \\
e^+e^-\rightarrow c\bar{c}(^3S_1^{[8]})+c\bar{c}(^3P_J^{[8]})+g, \label{eqn:process2a}
\eea
in addition to which, five processes are also at this order, as listed below, and will be calculated in our paper.
\bea
e^+e^-\rightarrow c\overline{c}(^3S_1^{[1]})+ c\overline{c}(^3S_1^{[8]})+g, \label{eqn:process2b1} \\
e^+e^-\rightarrow c\overline{c}(^1S_0^{[8]})+ c\overline{c}(^1S_0^{[8]})+g, \label{eqn:process2b2} \\
e^+e^-\rightarrow c\overline{c}(^3S_1^{[8]})+ c\overline{c}(^3S_1^{[8]})+g, \label{eqn:process2b3} \\
e^+e^-\rightarrow c\overline{c}(^1S_0^{[8]})+ c\overline{c}(^3P_J^{[8]})+g, \label{eqn:process2b4} \\
e^+e^-\rightarrow c\overline{c}(^3P_J^{[8]})+ c\overline{c}(^3P_J^{[8]})+g. \label{eqn:process2b5}
\eea
The representatvie Feynman diagrams for the final-state-gluon-emission processes are presented in Fig.(\ref{diagram-g}, \ref{diagram-h}, and \ref{diagram-i}).
However, not all the processes have the three types of diagrams.
So, we summarize the possible diagrams for each of the processes in Table\ref{tab:diag}.

\begin{table}[htbp]
\begin{center}
\caption{
Possible Feynman diagrams for each of the final-state-gluon-emission processes.
The processes are abbreviated to the numbers of the equations,
namely Eq.(\ref{eqn:process2a}, \ref{eqn:process2b1}, \ref{eqn:process2b2}, \ref{eqn:process2b3}, \ref{eqn:process2b4}, \ref{eqn:process2b5}),
in which the processes are presented.
}
\begin{tabular}{ccccccc}
\hline
\hline
Process~&~\ref{eqn:process2a}~&~\ref{eqn:process2b1}~&~\ref{eqn:process2b2}~&~\ref{eqn:process2b3}~&~\ref{eqn:process2b4}~&~\ref{eqn:process2b5} \\
\hline
Feynman diagrams~&~\ref{diagram-g}\ref{diagram-h}\ref{diagram-i}~&~\ref{diagram-g}\ref{diagram-i}~&~\ref{diagram-g}\ref{diagram-h}~
&~\ref{diagram-g}\ref{diagram-i}~&~\ref{diagram-g}\ref{diagram-h}~&~\ref{diagram-g}\ref{diagram-h} \\
\hline
\hline
\end{tabular}
\label{tab:diag}
\end{center}
\end{table}

Before we present the numerical results, we first address the divergences rising from the processes listed above.
First of all, the LO processes are divergence free, and we denote their total cross sections as $\sigma^{LO}=\sigma^{LO}_{^3S_1^{[8]}+^1S_0^{[8]}}+\sigma^{LO}_{^3S_1^{[8]}+^3P_J^{[8]}}$.
The virtual corrections to $\sigma^{LO}$ contain both ultraviolet (UV) and infrared (IR) divergences.
The UV divergences can be eliminated through the renormalization precedure,
while the IR divergences will be canceled by those emerging in the real corrections,
the processes for which are presented in Eq.(\ref{eqn:process2a}).
We denote the renormalized virtual-correction total cross sections as $\sigma^V=\sigma^V_{^3S_1^{[8]}+^1S_0^{[8]}}+\sigma^V_{^3S_1^{[8]}+^3P_J^{[8]}}$,
and the real corrections to $\sigma^{LO}$ as $\sigma^R=\sigma^R_{^3S_1^{[8]}+^1S_0^{[8]}}+\sigma^R_{^3S_1^{[8]}+^3P_J^{[8]}}$.
The complete QCD NLO corrections to $\sigma^{LO}$ can be expressed as
\be
\sigma^{NLO}\equiv\sigma^V+\sigma^R=\sigma^{NLO}_{^3S_1^{[8]}+^1S_0^{[8]}}+\sigma^{NLO}_{^3S_1^{[8]}+^3P_J^{[8]}}, \label{eqn:csnlo}
\ee
where
\bea
\sigma^{NLO}_{^3S_1^{[8]}+^1S_0^{[8]}}=\sigma^V_{^3S_1^{[8]}+^1S_0^{[8]}}+\sigma^R_{^3S_1^{[8]}+^1S_0^{[8]}}, \NO \\
\sigma^{NLO}_{^3S_1^{[8]}+^3P_J^{[8]}}=\sigma^V_{^3S_1^{[8]}+^3P_J^{[8]}}+\sigma^R_{^3S_1^{[8]}+^3P_J^{[8]}}. \label{eqn:csnlo2}
\eea
Both $\sigma^{NLO}_{^3S_1^{[8]}+^1S_0^{[8]}}$ and $\sigma^{NLO}_{^3S_1^{[8]}+^3P_J^{[8]}}$ are divergence free.
The total cross sections for the $J/\psi$ pair production through $^3S_1^{[8]}+^1S_0^{[8]}$ and $^3S_1^{[8]}+^3P_J^{[8]}$ channels are the sum of their LO and NLO contributions.
\bea
\sigma_{^3S_1^{[8]}+^1S_0^{[8]}}=\sigma^{LO}_{^3S_1^{[8]}+^1S_0^{[8]}}+\sigma^{NLO}_{^3S_1^{[8]}+^1S_0^{[8]}}, \NO \\
\sigma_{^3S_1^{[8]}+^3P_J^{[8]}}=\sigma^{LO}_{^3S_1^{[8]}+^3P_J^{[8]}}+\sigma^{NLO}_{^3S_1^{[8]}+^3P_J^{[8]}}. \label{eqn:completecsdiv}
\eea
Note that we adopt the on-shell (OS) renormalization scheme for the renormalization of c-quark mass and the wave functions of the c-quark and gluon,
and modified-mininum-subtraction ($\overline{MS}$) scheme for that of the QCD coupling constant,
which are coincide with Ref.~\cite{Gong:2007db}.
The corresponding renormalization constants,
$Z_m^{OS}$ (for the c-quark mass), $Z_2^{OS}$ (for the c-quark wave function),
$Z_3^{OS}$ (for the gluon wave function), and $Z_g^{\overline{MS}}$ (for the QCD coupling constant), are
\bea
&&\delta Z^{OS}_{m}=-3C_{F}\frac{\alpha_{s}}{4\pi}[\frac{1}{\epsilon_{UV}}-\gamma_{E}+\ln\frac{4\pi\mu^{2}_{r}}{m^{2}_{c}}+\frac{4}{3}], \NO \\
&&\delta Z^{OS}_{2}=-C_{F}\frac{\alpha_{s}}{4\pi}[\frac{1}{\epsilon_{UV}}+\frac{2}{\epsilon_{IR}}-3\gamma_{E}+3\ln\frac{4\pi\mu^{2}_{r}}{m^{2}_{c}}+4], \NO \\
&&\delta Z^{OS}_{3}=\frac{\alpha_{s}}{4\pi}[(\beta'_{0}-2C_{A})(\frac{1}{\epsilon_{UV}}-\frac{1}{\epsilon_{IR}})-\frac{4}{3}T_F(\frac{1}{\epsilon_{UV}}-\gamma_{E}+\ln\frac{4\pi\mu^{2}_{r}}{m^{2}_{c}})], \NO \\
&&\delta Z^{\overline{MS}}_{g}=-\frac{\beta_{0}}{2}\frac{\alpha_{s}}{4\pi}[\frac{1}{\epsilon_{UV}}-\gamma_{E}+\ln(4\pi)],
\eea
where $\mu_r$ is the renormalization scale, $\gamma_E$ is Euler's constant,
$\beta_0=\frac{11}{3}C_A-\frac{4}{3}T_Fn_f$ is the one-loop coefficient of the QCD beta function, $n_f$ is the number of active quark flavors.
In SU(3)$_c$, color factors are given by $T_F=\frac{1}{2}$, $C_F=\frac{4}{3}$, $C_A=3$, 
and $\beta'_{0}\equiv\beta_{0}+(4/3)T_F=(11/3)C_A-(4/3)T_Fn_{lf}$, where $n_{lf}\equiv n_f-1 =3$ is the number of light quark flavors.
Actually, in the NLO total amplitude level, the terms proportion to $\delta Z^{OS}_{3}$ 
cancel each other; thus the result is independent of the renormalization scheme of the gluon field.

The cross sections for the processes listed in Eq.(\ref{eqn:process2b4} and \ref{eqn:process2b5}) also have divergences,
which, however, can be eliminated through the renormalization of the SDCs for them.
We take process \ref{eqn:process2b4} as an example.
The cancellation of its divergences requires the calculation of the NLO corrections to $\langle O^{J/\psi}(^3S_1^{[8]})\rangle$.
The bare LDME can be expressed as
\be
\langle O^{J/\psi}(^3S_1^{[8]})\rangle_{bare}=\langle O^{J/\psi}(^3S_1^{[8]})\rangle-\frac{\alpha_s}{\pi m_c^2}\frac{N_c^2-4}{N_c}
(\frac{1}{\epsilon_{IR}}-\frac{1}{\epsilon_{UV}})\langle O^{J/\psi}(^3P_0^{[8]})\rangle, \label{eqn:ldmenlo}
\ee
where $m_c$ is the c-quark mass, and $N_c=3$ for SU(3) gauge theory.
Here we adopt the $\mu_\Lambda$-cutoff renormalization scheme~\cite{Jia:2014jfa} to subtract the UV divergence.
By substituting the relation between the bare and renormalized LDMEs,
\bea
\langle O^{J/\psi}(^3S_1^{[8]})\rangle_{bare}&=&\langle O^{J/\psi}(^3S_1^{[8]})\rangle_{renorm}+\frac{\alpha_s}{\pi m_c^2}\frac{N_c^2-4}{N_c} \NO \\
&\times&(\frac{1}{\epsilon_{UV}}-\gamma_E+\frac{5}{3}+ln(\frac{\pi\mu^2}{\mu_\Lambda^2}))\langle O^{J/\psi}(^3P_0^{[8]})\rangle, \label{eqn:renormldme}
\eea
into Eq.(\ref{eqn:ldmenlo}), we obtain the renormalized LDME as
\bea
\langle O^{J/\psi}(^3S_1^{[8]})\rangle_{renorm}&=&\langle O^{J/\psi}(^3S_1^{[8]})\rangle-\frac{\alpha_s}{\pi m_c^2}\frac{N_c^2-4}{N_c} \NO \\
&\times&(\frac{1}{\epsilon_{IR}}-\gamma_E+\frac{5}{3}+ln(\frac{\pi\mu^2}{\mu_\Lambda^2}))\langle O^{J/\psi}(^3P_0^{[8]})\rangle. \label{eqn:ldmerenorm}
\eea
Then process $e^+e^-\rightarrow c\bar{c}(^3S_1^{[8]})+c\bar{c}(^1S_0^{[8]})$ contribute an additional divergent term
\bea
\sigma_{div}(e^+e^-\rightarrow c\bar{c}(^3S_1^{[8]})+c\bar{c}(^1S_0^{[8]}))=-\frac{\alpha_s}{\pi m_c^2}\frac{N_c^2-4}{N_c}(\frac{1}{\epsilon_{IR}}-\gamma_E+\frac{5}{3}+ln(\frac{\pi\mu^2}{\mu_\Lambda^2})) \NO \\
\times\hat{\sigma}(e^+e^-\rightarrow c\bar{c}(^3S_1^{[8]})+c\bar{c}(^1S_0^{[8]}))\langle O^{J/\psi}(^3P_0^{[8]})\rangle, \label{eqn:sigmadiv}
\eea
which cancels the IR sigularities rising from process \ref{eqn:process2b4}.
In this sense, we can redefine the SDC for process \ref{eqn:process2b4} as
\bea
\hat{\sigma}_{renorm}(e^+e^-\rightarrow c\bar{c}(^1S_0^{[8]})+c\bar{c}(^3P_J^{[8]})+g)=\hat{\sigma}(e^+e^-\rightarrow c\bar{c}(^1S_0^{[8]})+c\bar{c}(^3P_J^{[8]})+g) \NO \\
-\frac{\alpha_s}{\pi m_c^2}\frac{N_c^2-4}{N_c}(\frac{1}{\epsilon_{IR}}-\gamma_E+\frac{5}{3}+ln(\frac{\pi\mu^2}{\mu_\Lambda^2}))\hat{\sigma}(e^+e^-\rightarrow c\bar{c}(^3S_1^{[8]})+c\bar{c}(^1S_0^{[8]})), \label{eqn:sdcfinite}
\eea
where
\bea
\hat{\sigma}(e^+e^-\rightarrow c\bar{c}(^1S_0^{[8]})+c\bar{c}(^3P_J^{[8]})+g)=\hat{\sigma}(e^+e^-\rightarrow c\bar{c}(^1S_0^{[8]})+c\bar{c}(^3P_0^{[8]})+g) \NO \\
+3\hat{\sigma}(e^+e^-\rightarrow c\bar{c}(^1S_0^{[8]})+c\bar{c}(^3P_1^{[8]})+g)+5\hat{\sigma}(e^+e^-\rightarrow c\bar{c}(^1S_0^{[8]})+c\bar{c}(^3P_2^{[8]})+g) \label{eqn:def3pj8}
\eea
has been implicated in Eq.(\ref{eqn:sdcfinite}) (the same convention applies to the SDCs with the subscript "renorm").
$\hat{\sigma}_{renorm}$ is a finite quantity, therefore, we can replace, in Eq.(\ref{eqn:total}), the divergent one by it.
The same operation can be done for process \ref{eqn:process2b5} as well.
Then, we denote all the divergence-free total cross sections for the processes listed in
Eq.(\ref{eqn:process2b1}, \ref{eqn:process2b2}, \ref{eqn:process2b3}, \ref{eqn:process2b4}, \ref{eqn:process2b5}) as $\sigma_{n_1+n_2}$,
where $n_1$ and $n_2$ are the corresponding $c\bar{c}$ states.

\section{Numerical results}\label{cha:3}

In our analytic calculation,
we use our {\it Mathematica} package with the employment of FeynArts~\cite{Hahn:2000kx}, FeynCalc~\cite{Mertig:1990an}, FIRE~\cite{Smirnov:2008iw} and Apart~\cite{Feng:2012iq}.
As a cross check, we also compute the processes using the FDC package~\cite{Wang:2004du}, except for process \ref{eqn:process2b5}.
To subtract the IR divergences in the gluon-emission processes, we adopt the two-cutoff slicing strategy~\cite{Harris:2001sx}.
The independence of the cutoff has been checked.

We have the following global choices of the parameters in our calculation:
$\alpha=1/137$, and the colliding energy of the electron and positron is $\sqrt{s}=10.6\gev$.
The $J/\psi$ mass is fixed to $M_{J/\psi}=2m_c$ to keep the gauge invariance.
The default values of $m_c$ and the renormalization scale ($\mu_r$) are $m_c=1.5\gev$ and $\mu_r=3.0\gev$, respectively.
Since we investigate the $\mu_r$ dependence of the total cross sections,
the two-loop running $\alpha_s$ is employed in our computation.
The values of the SDCs for all the processes are listed in Table \ref{tab:sdc},
where the SDCs for the $^3P_J^{[8]}$ channels are defined by multiplying a factor of $m_c^2$ to those defined in Ref.~\cite{Bodwin:1994jh},
in order to keep the homogeneity of the dimensions (for double $^3P_J^{[8]}$ production, this factor should be $m_c^4$).
The LO SDCs for $^1S_0^{[8]}+^3S_1^{[8]}$ and $^3S_1^{[8]}+^3P_J^{[8]}$ productions are $60.07\fb/\gev^3$ and $290.9\fb/\gev^3$, respectively.

\begin{table}[htbp]
\begin{center}
\caption{
The values of $\hat{\sigma}_{n_1+n_2}$ in ($\fb/\gev^3$) up to order $\mathcal{O}(\alpha^2\alpha_s^3)$.
The convention introduced in Eq.(\ref{eqn:def3pj8}) is adopted.
}
\begin{tabular}{ccccc}
\hline
\hline
$n_2$ \slash $n_1$~&~$^3S_1^{[1]}$~&~$^1S_0^{[8]}$~&~$^3S_1^{[8]}$~&~$^3P_J^{[8]}$~ \\
\hline
$^3S_1^{[1]}$~&~0~&~0~&~4.55~&~0 \\
\hline
$^1S_0^{[8]}$~&~0~&~ 0.28~&~ 167.48~&~-74.37 \\
\hline
$^3S_1^{[8]}$~&~ 4.55~&~ 167.48~&~ 2.48 ~&~ 815.77 \\
\hline
$^3P_J^{[8]}$~&~0~&~ -74.37~&~ 815.77~&~ -265.29 \\
\hline
\hline
\end{tabular}
\label{tab:sdc}
\end{center}
\end{table}

Employing the LDMEs obtained in Refs.~\cite{Zhang:2014ybe, Sun:2015pia},
namely
\bea
\langle O^{J/\psi}(^3S_1^{[1]})\rangle&=&0.65 \gev^3 \NO \\
\langle O^{J/\psi}(^1S_0^{[8]})\rangle&=&0.78\times 10^{-2} \gev^3 \NO \\
\langle O^{J/\psi}(^3S_1^{[8]})\rangle&=&1.08\times 10^{-2} \gev^3 \NO \\
\langle O^{J/\psi}(^3P_0^{[8]})\rangle /m_c^2 &=&2.01\times 10^{-2} \gev^3 \label{eqn:ldmevalue}
\eea
we list the cross sections for each channel in Table \ref{tab:cs}.

\begin{table}[htbp]
\begin{center}
\caption{
The values of $\sigma_{n_1+n_2}$ in ($\fb$) up to order $\mathcal{O}(\alpha^2\alpha_s^3)$.
The convention introduced in Eq.(\ref{eqn:def3pj8}) is adopted.
The LDMEs are taken from Refs.~\cite{Zhang:2014ybe, Sun:2015pia}.
}
\begin{tabular}{ccccc}
\hline
\hline
$n_2$ \slash $n_1$~&~$^3S_1^{[1]}$~&~$^1S_0^{[8]}$~&~$^3S_1^{[8]}$~&~$^3P_J^{[8]}$~ \\
\hline
$^3S_1^{[1]}$~&~0~&~0~&~ 0.032~&~0 \\
\hline
$^1S_0^{[8]}$~&~0~&~1.68$\times10^{-5}$~&~ 0.014~&~ -0.012 \\
\hline
$^3S_1^{[8]}$~&~ 0.032~&~ 0.014~&~2.90$\times10^{-4}$~&~ 0.177 \\
\hline
$^3P_J^{[8]}$~&~0~&~ -0.012~&~0.177~&~-0.107 \\
\hline
\hline
\end{tabular}
\label{tab:cs}
\end{center}
\end{table}

The total cross section for double $J/\psi$ production at B factories up to order $\mathcal{O}(\alpha^2\alpha_s^3)$ is the sum of those for different channels.
Note that the two processes which are symmetric in the sense of switching $n_1$ and $n_2$ are only counted once to avoid the double counting.
We obtain this value as $\sigma=0.1046\gev$.
Compared with the results obtained in Ref.~\cite{Gong:2008ce},
it is even smaller than that via the CS channels up to QCD NLO.

To investigate the uncertainties brought in by the two scales,
we vary $m_c$ from 1.2$\gev$ to 1.7$\gev$ and $\mu_r$ from $3.0\gev$ to $\sqrt{s}/2$ and calculate the corresponding total cross sections.
When one of these scales varies its value, the other is fixed.
Note that the LDMEs used in our calculation are obtained with the configuration $m_c=1.5\gev$.
When investigating the $m_c$ dependence, we need to take the scaling rule, $\langle O^{J/\psi}(n)\rangle\propto m_c^3$, into account.

\begin{figure}[!ht]
 \begin{center}
 \includegraphics[width=8.0cm]{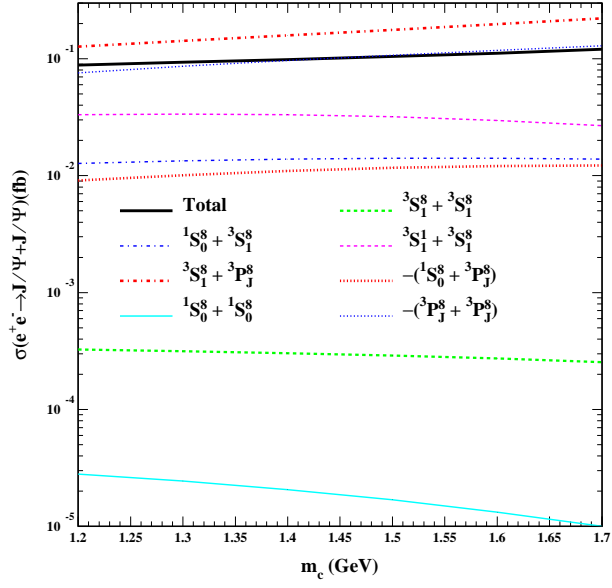}
 \end{center}
 \caption{$\sigma$ as a function of $m_c$. The renormalization scale is fixed to $\mu_r=3.0\gev$.}
 \label{fig:fmc}
\end{figure}

\begin{figure}[!ht]
 \begin{center}
 \includegraphics[width=8.0cm]{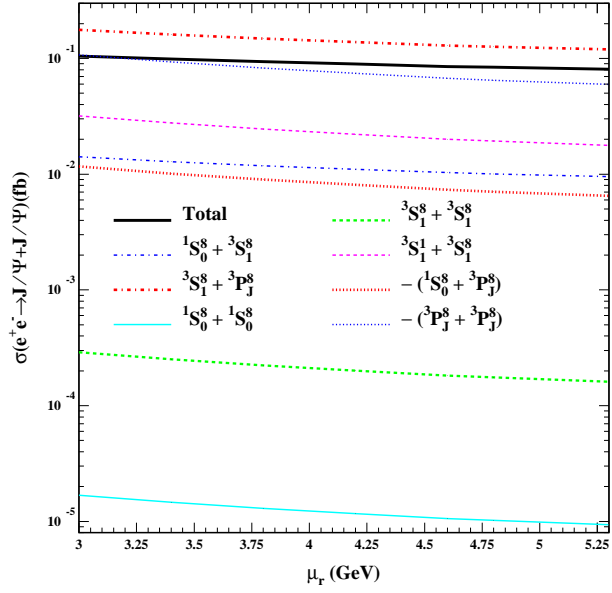}
 \end{center}
 \caption{$\sigma$ as a function of $\mu_r$. The c-quark mass is fixed to $m_c=1.5\gev$.}
 \label{fig:miu}
\end{figure}

The total cross section as a function of charm quark mass $m_c$ is presented in Fig.\ref{fig:fmc}.
We can see that the $^3S_1^{[8]}+^3P_J^{[8]}$ and $^3P_J^{[8]}+^3P_J^{[8]}$ channels provide the largest contributions,
while the others contribute smaller with visible hierachy. 
Expecially, both the $^1S_0^{[8]}+^3P_J^{[8]}+g$ and $^3P_J^{[8]}+^3P_J^{[8]} + g$ cross sections are negative .
The total cross section increases from about 0.08 $\fb$ to about 0.12 fb as the $m_c$ increases from $1.2\gev$ to $1.7\gev$.
The $\mu_r$ dependence of the total cross section is presented in Fig.\ref{fig:miu},
and the $\sigma$ decreases from 0.104$\fb$ to 0.08$\fb$ as $\mu_r$ increases from $3.0\gev$ to $\sqrt{s}/2$.
The dependence on the two scales is not severe, which indicates good convergence of the perturbative expansion.

\begin{table}[htbp]
\begin{center}
\caption{
\label{tab:ldme}
The cross sections of double $J/\psi$ production and their corresponding $J/\psi$ LDMEs values from Ref.\cite{Butenschoen:2011yh,Chao:2012iv,Gong:2012ug,Bodwin:2014gia}.
}
\begin{tabular}{cccccc}
\hline
\hline
Ref.&Butenschon,&Chao, Ma, Shao,&Gong, Wan, Wang,&Bodwin, Chung, \\
&Kniehl~\cite{Butenschoen:2011yh}&Wang, Zhang~\cite{Chao:2012iv}&Zhang~\cite{Gong:2012ug}&Kim, Lee~\cite{Bodwin:2014gia} \\
\hline
$\langle{\cal O}^{H}(^{3}S^{[1]}_{1})\rangle$(GeV$^3$)~&1.32~&1.16~&1.16~&   \\
$\langle{\cal O}^{H}(^{1}S^{[8]}_{0})\rangle$(GeV$^3$)~&3.04$\times10^{-2}$~& 8.9$\times10^{-2}$~&9.7$\times10^{-2}$&9.9$\times10^{-2}$  \\
$\langle{\cal O}^{H}(^{3}S^{[8]}_{1})\rangle$(GeV$^3$)~&1.6$\times10^{-3}$~& 3.0$\times10^{-3}$~&-4.6$\times10^{-3}$&1.1$\times10^{-2}$ &   \\
$\langle{\cal O}^{H}(^{3}P^{[8]}_{0})\rangle$(GeV$^5$)~&-9.1$\times10^{-3}$~& 1.26$\times10^{-2}$~&-2.14$\times10^{-2}$& 1.1$\times10^{-2}$ \\
\hline
$\sigma(J/\psi)$(fb)& 0.018&~ 0.031&~ -0.016 &~ 0.245
\footnote[1]{Since the CS LDME was not given in Ref.~\cite{Bodwin:2014gia}, we adopt the most frequently used value $\langle{\cal O}^{H}(^{3}S^{[1]}_{1})\rangle$=1.16 GeV$^3$ in the calculation.}  \\
\hline
\hline
\end{tabular}
\end{center}
\end{table}

Since there are several parallel extractions of the LDMEs, we need to investigate the uncertainties brought in by the different values of them.
As is shown in TABLE\ref{tab:ldme}, the total cross sections obtained by using the LDMEs in Ref.~\cite{Bodwin:2014gia} are almost twice of ours,
however, still too small to be observed by the experiment.

\section{Summary and conclusion}\label{cha:4}

We calculated the total cross sections for double $J/\psi$ production in $e^+e^-$ annihilation at the B-factory energy up to $\mathcal{O}(\alpha^2\alpha_s^3)$ within the framework of NRQCD.
We studied the $m_c$ and $\mu_r$ dependence of the total cross sections,
and found that the results ranges from $0.08\fb$ to $0.12\fb$.
Also, we investigated the uncertainties by trying different set of the LDMEs.
Even for the largest results, the total cross section is too small for Belle to observe any significant access.
This result is consistent with the Belle measurement.

\acknowledgments
{This work is supported by the National Nature Science Foundation of China (No. 11405268).}


\end{document}